\documentclass[%
 reprint,
nofootinbib,
 amsmath,amssymb,
 aps,
]{revtex4-2}

\pdfoutput=1 

\usepackage{graphicx}
\usepackage{braket} 
\usepackage{dcolumn}
\usepackage{bm}
\usepackage{appendix}

\begin{document}

\preprint{APS/123-QED}

\title{Copycat process in the early stages of einselection}

\author{Rose Baunach}
 \email{baunach@ucdavis.edu}
 \author{Andreas Albrecht}%
 \email{ajalbrecht@ucdavis.edu}
\affiliation{Center for Quantum Mathematics and Physics and Department of Physics and Astronomy\\ UC Davis, One Shields Ave, Davis CA.}

\author{Andrew Arrasmith}%
 \email{aarrasmith@lanl.gov}
\affiliation{Theoretical Division, Los Alamos National Laboratory, Los Alamos, NM USA.}

\medskip

\date{\today}

\begin{abstract}
We identify and describe unique early time behavior of a quantum system initially in a superposition, interacting with its environment.  This behavior---the copycat process---occurs after the system begins to decohere, but before complete einselection. To illustrate this behavior analytic solutions for the system density matrix, its eigenvalues, and eigenstates a short time after system-environment interactions begin are provided. Features of the solutions and their connection to observables are discussed, including predictions for the continued evolution of the eigenstates towards einselection, time dependence of spin expectation values, and an estimate of the system's decoherence time. In particular we explore which aspects of the early stages of decoherence exhibit quadratic evolution to leading order, and which aspects exhibit more rapid linear behavior. Many features of our early time perturbative solutions are agnostic of the spectrum of the environment. We also extend our work beyond short time perturbation theory to compare with numerical work from a companion paper.

\end{abstract}

\maketitle


\section{\label{sec:intro}Introduction}

There are many reasons why one would want to study the effects of decoherence and einselection on a quantum system interacting with its environment---from interest in theoretical interpretations of quantum mechanics to applications in quantum computing \cite{Schlosshauer,Zurek:2003zz,JoosZeh,Preskill,Unruh:1994az,CPZ1,BreuerPetruccione,MikeNIke}. In~\cite{ACLintro} we introduced the ``adapted Caldeira Leggett'' (ACL) model, a tool designed to explore these phenomena using fully unitary calculations in the combined system-environment space.  This tool enables us to examine behaviors outside of the standard approximation schemes common in the field.  

\begin{figure}[h!]
\centering
\includegraphics[width=3.4in]{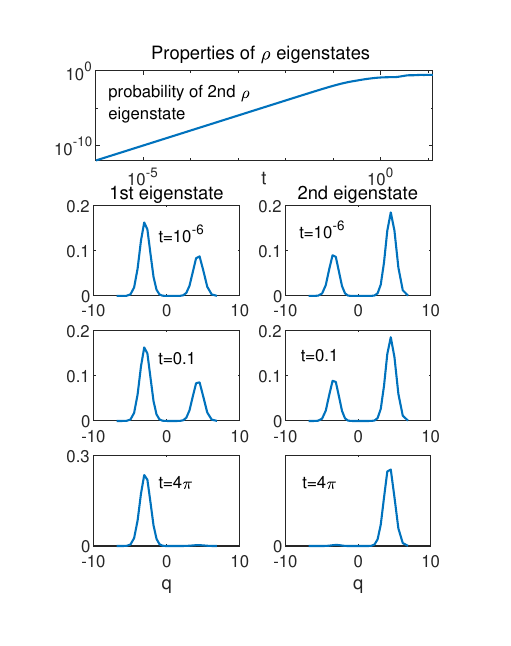}
\vspace{-1.0cm}
\caption{Copycats in the early stages of entanglement: A system that is initially in a pure superposition of two coherent states becomes entangled with the environment as it evolves.  We plot the second eigenvalue and eigenstates of the system density matrix $\rho$ for early times.  The 2nd eigenstate takes the mirror image ``copycat'' form over several decades of evolution. The bottom row shows these two states after einselection is complete, and the initial superposition has become a mixture of classical wave packets.  The eigenstate plots show $\left|\psi(q)\right|^2$ for the first two eigenstates, where $q$ is a generalized position. Time is shown in units where the oscillator period is $2\pi$. (The absence of phase information means the orthogonality of the eigenstates does not appear manifest in these plots.)
}
\label{fig:CopyCat}
\end{figure}

Our original aim was a study of the relationship between einselection and the arrow of time.  We present the outcome of that work in~\cite{ACLeqm}.  On the path of exploring decoherence and einselection with the ACL model numerically, we witnessed a curious phenomenon---the copycat process---which we investigate in this paper.
Figure~\ref{fig:CopyCat} (Fig.~12 in~\cite{ACLintro}) gives a general picture of this process. 

The ACL model describes a simple harmonic oscillator (SHO) coupled to an environment.  For Fig.~\ref{fig:CopyCat} the SHO was set up in a ``Schr\"odinger cat'' superposition of coherent state classical wavepackets.  Over time, decoherence with the environment brings the SHO into a classical mixture of wavepackets described by a density matrix ($\rho$) with the wavepackets as eigenstates.  In this manner the classical wavepackets are specially selected by the specifics of the decoherence physics, a process called ``einselection.'' 

The evolution shown in Fig.~\ref{fig:CopyCat} starts with the system and environment in a product state.  At that moment the SHO density matrix, $\rho$, has only one nonzero eigenvalue.  An instant later a second nonzero (but infinitesimal) eigenvalue emerges. As soon as this second eigenvalue becomes resolved in our calculations, the corresponding eigenstate takes on the intriguing ``mirror image'' (copycat) form shown in Fig.~\ref{fig:CopyCat}. The first and second eigenstates keep these forms as the second eigenvalue evolves over many decades in magnitude.  Eventually on a timescale given by the decoherence time, einselection takes place.~\footnote{The SHO has a period of $2\pi$. Thus, choosing $t=4\pi$ for the final time shown in Fig.~\ref{fig:CopyCat} simplifies our presentation. More details about the behavior of the ACL model can be found in~\cite{ACLintro}.}~\footnote{In Fig.~\ref{fig:CopyCat}, $q$ labels the discrete set of eigenvalues of the (dimensionless) position operator of the ACL model~\cite{ACLintro}, and $\psi(q)$ used here corresponds to $\psi_{\alpha}(q)$ (from Eqn.~4 in~\cite{ACLintro}), the discrete set of coefficients from expanding the state in eigenstates of position. In the panels giving eigenstates the discrete set of points $\left|\psi(q)\right|^2$ are connected to guide the eye, but there is no actual continuum and $\left|\psi(q)\right|^2$ is dimensionless. }

While initially the copycat phenomenon seemed striking and unusual to us, and we were a bit concerned about the possible role of numerical artifacts in our results, we have come to understand this process in relatively simple physical terms.  The ``transient stability'' we observe relates to the slow quadratic start to the early time evolution of the eigenstates. 
Furthermore, the mirror image copycat form seems quite natural if one thinks of the two wavepackets as spanning a two dimensional reduced Hilbert space (effectively a single qubit). From that point of view, the form shown is the only option for the 2nd eigenstate of $\rho$ until the first eigenstate has time to evolve appreciably.  In fact, the copycat picture we describe here is implicit in standard results from Nuclear Magnetic Resonance (NMR) physics, although they have not been previously presented from that point of view. 

This work originally came about through our need to fully understand the behaviors of the ACL model before applying it to the new problems explored in~\cite{ACLeqm}. Having satisfied ourselves that the copycat process was indeed physical, and having extended it in a number of directions, we've decided those explorations are worth reporting in this paper.  We feel they may be of interest to those studying the early stages of decoherence, perhaps in the context of quantum technologies.  In particular we offer explorations of the extent to which the early stages of decoherence are quadratic to leading order, and which aspects evolve with a more rapid linear behavior. 

 The rest of this paper is devoted to describing and analytically deriving the copycat process. In Sect.~\ref{sec:formalism} we briefly review the ACL and reduced Caldeira-Leggett (RCL) models as presented in \cite{ACLintro}. Then in Sect.~\ref{sec:CCintro} we present perturbative analysis of the copycat process by calculating the system density matrix, associated eigenvalues, and eigenstates of a two-state system entangled with its environment a short time after system-environment interactions begin. In our derivations we chose to model the superposition of coherent states in Eqn.~\ref{eq:3} as a two state system, both for simplicity and because such coherent superpositions were observed to behave as essentially a two-state system in our numerical work \cite{ACLintro}. Section~\ref{sec:Implications} is a discussion of some specific features and applications of our derived solutions for the two state system, including predictions for the continued evolution of the system eigenstates, early time behavior of spin observables, and an estimate of the system's decoherence time.   In Sect.~\ref{sec:Beyond} we extend our solutions beyond short time perturbation theory to compare with our numerical work in \cite{ACLintro}. We also comment on the duration of the perturbative copycat regime in Sect.~\ref{sec:Numerical} and graphically demonstrate that the perturbative solutions can model the full numerical evolution for a sizable span of time. Our further discussion and conclusions are provided in Sect.~\ref{sec:conclude}, which include discussion of how the copycat process generalizes for initial states with larger numbers of `cats' and comparisons of our work with existing literature.  
 
 Appendix~\ref{sec:qtrit} extends our technical results from a qubit model to the qutrit case, providing additional perspective on the possibility of linear behaviors (which we summarize is Sect.~\ref{sec:conclude}). These investigations offer further nuance to the technical aspects of the copycat process discussed in the main text, as we discover the possibility of linear time evolution in the orthogonal eigenstates for some cases, even though the evolution of the eigenvalues and initial state remains quadratic, as in the qubit case.  We also illustrate how results from our qutrit analysis are reflected in the behaviors of more complicated ACL-like models with three wavepacket Schr\"odinger cat initial states. 


\section{\label{sec:formalism}Formalism}

\subsection{\label{sec:basics}Basics}
We consider a ``world'' Hilbert space, comprised of a system and an environment: $w = s \otimes e$. We take $\ket{\psi_{w}}$ initially to be a product state and study the entanglement caused by  system-environment interactions:
\begin{equation} \label{eq:1}
\ket{\psi_{w}} = \ket{\psi_{s}}  \ket{\psi_{e}} \quad \overrightarrow{entanglement} \quad  \sum_{i,j} b_{ij} \ket{i}_{s} \ket{j}_{e}.
\end{equation}
The onset of system-environment entanglement is called decoherence. 
Once entanglement has taken place, the system is described by the density matrix
\begin{equation} \label{eq:2}
    \rho_s = Tr_{e}(\ket{\psi_{w}} \bra{\psi_{w}}).
\end{equation}
Einselection is the special case of decoherence where environmental interactions induce the system density matrix to become diagonal in a preferred basis.  These preferred basis states are called pointer states in the literature \cite{Zurek1982,Zurek:2003zz}.

Under certain conditions (which apply in Fig.~\ref{fig:CopyCat}) coherent states will be einselected as the pointer states \cite{Coherent,Schlosshauer,ACLintro}.  Under those conditions, a system that starts in a superposition of coherent states 
\begin{equation} \label{eq:3}
\ket{\psi_{s}} = a_{1}\ket{\alpha_{1}} + a_{2}\ket{\alpha_{2}} 
\end{equation}
would evolve into the density matrix
\begin{equation}  \label{eqn:einselect}
    \rho_s = |a_1|^{2}\ket{\alpha_1}\bra{\alpha_1} + |a_2|^{2}\ket{\alpha_2}\bra{\alpha_2}. 
\end{equation}
Here we have labeled coherent states with the parameter $\alpha$, a standard convention articulated in detail for the ACL model in~\cite{ACLintro}.  Indeed Eqn.~\ref{eqn:einselect} roughly describes what we see in Fig.~\ref{fig:CopyCat}.  (Inspection of the full analysis presented in~\cite{ACLintro} reveals that in this particular case, the finite form of the ACL model leads to a number of small deviations from the idealized picture described by Eqn.~\ref{eqn:einselect}.) 
It is in the early stages of the evolution toward the Eqn.~\ref{eqn:einselect} form that we notice the copycat behavior.   

As illustrated in Fig.~\ref{fig:CopyCat}, at early times there is an eigenstate of the system density matrix that resembles the initial state and one that has the ``copycat'' form.  The initial evolution of these eigenstates are quadratic in time---as we will prove subsequently---or ``slow,'' hence the appearance of  ``transient stability.'' We call the appearance of a copycat state and its subsequent behavior the ``copycat process".  To the best of our knowledge the copycat process and its implications have not been directly explored in the literature, although we will discuss how earlier work has come very close to this topic in an indirect way.  

\vspace{-0.6cm}
\subsection{\label{sec:ACL} The ACL Model}
\vspace{-0.2cm}
As discussed in \cite{ACLintro}, the original Caldeira Leggett (CL) model is a toy model describing a system interacting with its environment with a Hamiltonian of the form \cite{Caldeira:1982iu,Schlosshauer,BreuerPetruccione}:
\vspace{-0.2cm}
\begin{equation} \label{eq:5}
 H_{w}= H^{s}_{SHO} \otimes 1^{e} + q_{SHO} \otimes H^{e}_{I} + H^{e} \otimes 1^{s}.
 \end{equation}
 In the CL model the system is a simple harmonic oscillator (SHO) moving in the standard SHO potential, and the environment is an infinite set of SHOs. The system and environment together describe a closed system undergoing unitary evolution, but generally the system and environment individually do not have to evolve unitarily.  
 
 The adapted Caldeira Leggett (ACL) model was introduced as an adaptation of the CL model which operates in a finite dimensional Hilbert space---so its evolution can be investigated numerically in its full unitary form \cite{ACLintro}. In the ACL model, the Hamiltonian is also given by Eqn.~\ref{eq:5}, but the components are modified since the Hilbert space is finite---for example, the system is given by a truncated SHO. Full technical details are given in \cite{ACLintro}.  
 

\vspace{-0.2cm}
\section{\label{sec:CCintro} Modelling the Copycat Process}

\subsection{\label{sec:setup}Setting up and the RCL model}
We start our technical explorations of the copycat process with the following observation: The ACL model used to produce Fig.~\ref{fig:CopyCat} had parameters adjusted to make the coherent states especially stable, making them the pointer states.  We also note that while the coherent state wavefunctions are nowhere truly zero, the overlap between the two coherent states shown in Fig.~\ref{fig:CopyCat} is exponentially suppressed making the two coherent states essentially orthogonal.  The SHO dynamics will ultimately move the two packets into positions of greater overlap, but the copycat process takes place on time scales short compared to the SHO evolution.  We use both the stability of the coherent states and their lack of overlap to argue heuristically that they span a two dimensional subspace, which is effectively decoupled from the rest of the SHO Hilbert space at early times. Based on these considerations, we model the superposition of coherent states in Eqn.~\ref{eq:3} (used to generate Fig.~\ref{fig:CopyCat}) with a single qubit coupled to an environment and undertake analytical calculations of early time behavior using perturbation theory in the small time parameter. 


As with the full SHO case, we start with a pure product state at $t=0$ with no initial entanglement. At $t=0$ the system (now just a qubit) is a two-state superposition and the environment is in some pure state which we call $\ket{\phi_{e}}$:
\begin{eqnarray} \label{eq:6}
 \ket{\psi_{w} (0)} & = & \ket{\psi_s(0)}\ket{\phi_{e}} \nonumber \\ 
  & = & \big(a\ket{\uparrow} + b\ket{\downarrow}\big) \ket{\phi_{e}}.
 \end{eqnarray}
Here $a$ and $b$ can be complex, and $ |a|^{2} + |b|^{2} = 1. $
We consider a Hamiltonian given by:
\begin{equation} \label{eqn:Hrcl}
H_{w} \equiv H = \lambda \big(\ket{\uparrow}\bra{\uparrow} H^{\uparrow}_{e} + \ket{\downarrow}\bra{\downarrow} H^{\downarrow}_{e} \big)
\end{equation}
where $\lambda$ is a real parameter to adjust the strength of the interaction, and $H^{\uparrow}_{e}$ and $H^{\downarrow}_{e}$ only operate in the subspace of the environment.  We will refer to $H_{w}$ as $H$ for brevity in what follows. 

We take 
$H^{\uparrow}_{e}$, $H^{\downarrow}_{e}$, and $H$ to all be time independent, and we generally allow $H^{\uparrow}_{e}H^{\downarrow}_{e} \neq H^{\downarrow}_{e}H^{\uparrow}_{e}.$  For most of what follows, no additional assumptions are made about the eigenvalue spectra of the $H_e$'s or the dimensionality of the environment.  We note that $H$ in Eqn.~\ref{eqn:Hrcl} is very similar to what we call the ``reduced Caldeira-Leggett'' (RCL) model Hamiltonian in~\cite{ACLintro}, although there we considered the special case where $H^{\uparrow}_{e} = -H^{\downarrow}_{e}$.  We call the model we use here an RCL model as well, and note that as discussed in~\cite{ACLintro} this model will einselect the pointer states $\ket{\uparrow}$ and $\ket{\downarrow}$.

Working in the Schrodinger picture with:
\begin{equation} \label{eq:8}
\imath \hbar \frac{\partial}{\partial t}\ket{\psi} = H\ket{\psi}  \quad \quad \quad U(t) = e^{\frac{- \imath H t}{\hbar}}
\end{equation}
we can perturbatively compute the state of the system and environment at a short time $t=\Delta$, using the series expansion of the time evolution operator
\begin{align} \label{eq:9}
    \ket{\psi_{w} (\Delta)} & = U(\Delta) \ket{\psi_{w} (0)} \notag \\
    {} & = \Big(1 - \frac{\imath H \Delta}{\hbar} - \frac{1}{2} \frac{(H \Delta)^{2}}{\hbar^{2}} + O(\Delta^{3})\Big) \ket{\psi_{w} (0)}.
\end{align}
The result is: 
\begin{align} \label{eq:10}
\begin{split}
    \ket{\psi_{w} (\Delta)} & =  \big(a\ket{\uparrow} + b\ket{\downarrow}\big)\ket{\phi_{e}}\\
        &  \hspace{0.5cm} - \frac{\imath \Delta \lambda}{\hbar}\big(a\ket{\uparrow} H^{\uparrow}_{e}\ket{\phi_{e}} + b\ket{\downarrow} H^{\downarrow}_{e}\ket{\phi_{e}}\big)\\
        &  \hspace{0.5cm}  - \frac{\Delta^{2} \lambda^{2}}{2\hbar^{2}} \big(a\ket{\uparrow} H^{\uparrow}_{e} H^{\uparrow}_{e} \ket{\phi_{e}} + b\ket{\downarrow} H^{\downarrow}_{e} H^{\downarrow}_{e} \ket{\phi_{e}}\big).
\end{split}
\end{align}
We have found the important leading order behavior occurs at second order, so we keep terms up to  $O(\Delta^{2})$ in what follows.

\subsection{\label{sec:level2} System Reduced Density Matrix}
We compute the reduced density matrix of the system after a short time $t = \Delta$:
\begin{equation} \label{eq:11}
 \rho_{s} (t = \Delta) = \mathbf{Tr_{e}}\big(\ket{\psi_{w} (\Delta)} \bra{\psi_{w} (\Delta)}\big).
 \end{equation}
The result expressed in the $\ket{\uparrow},\ket{\downarrow}$ basis is:
\begin{equation} \label{eq:12}
 \rho_{s}(\Delta) = \begin{bmatrix} a a^{*} & a b^{*} \Big( 1 + \imath \beta \Delta - \eta \Delta^{2} \Big) \\
 b a^{*} \Big( 1 - \imath \beta \Delta - \eta^{*} \Delta^{2} \Big) & b b^{*}
 \end{bmatrix}
\end{equation}
where the coefficients $\beta$ and $\eta$ are given by:
\begin{align}    
\beta & =  \frac{\lambda}{\hbar}\big(\bra{\phi_{e}}  H^{\downarrow}_{e}\ket{\phi_{e}} - \bra{\phi_{e}} H^{\uparrow}_{e} \ket{\phi_{e}} \big) \label{eq:13}\\
\eta &  =  \frac{\lambda^{2}}{\hbar^{2}} \Big( \frac{\bra{\phi_{e}}H^{\uparrow}_{e}H^{\uparrow}_{e}\ket{\phi_{e}} + \bra{\phi_{e}}H^{\downarrow}_{e}H^{\downarrow}_{e}\ket{\phi_{e}}}{2} \notag \\
& \qquad \quad  - \bra{\phi_{e}} H^{\downarrow}_{e}H^{\uparrow}_{e} \ket{\phi_{e}} \Big). \label{eq:14}
\end{align} 
To obtain the above it is necessary to recognize that $ \bra{\phi_{e}}H^{\downarrow}_{e}\ket{\phi_{e}}$ and  $\bra{\phi_{e}} H^{\uparrow}_{e}\ket{\phi_{e}}$ are real numbers, but that $\bra{\phi_{e}} H^{\uparrow}_{e} H^{\downarrow}_{e}\ket{\phi_{e}}$ can be complex (since $H^{\uparrow}_{e}H^{\downarrow}_{e} \neq H^{\downarrow}_{e}H^{\uparrow}_{e}$).  This requires $\beta$ to be purely real, but allows $\eta$ to be complex. It follows that the system density matrix in Eqn.~\ref{eq:12} is Hermitian and properly normalized since $\rho^{\dagger} = \rho$, and $Tr[\rho] = |a|^{2} + |b|^{2} = 1$. 
\vspace{0.4cm}

\subsection{\label{sec:2by2} Eigenvalues and eigenvectors of $\rho_s$}
\vspace{-0.2cm}
We use the general analytic form for the eigenvalues and eigenstates of a 2 x 2 Hermitian matrix. After obtaining the exact solutions from Eqn.~\ref{eq:12}, we then compute the series expansions in $\Delta$---keeping terms to $O(\Delta^2)$---to obtain the following perturbative expressions:
\begin{widetext}
\begin{align}
\ket{\psi_{1}} & = \Big( \frac{b^{*}}{|b|} \Big) \Bigg[  a \Bigg( 1 + \imath \beta \Delta + \Delta^{2}\big[ \epsilon \big( \frac{2 |a|^{2}|b|^{2} + |a|^{2}}{2}\big) - \eta \big] \Bigg) \ket{\uparrow} + b \Bigg(1 + \Delta^{2} \epsilon\Big(\frac{2|a|^{2}|b|^{2} - |a|^{2}}{2} \Big) \Bigg) \ket{\downarrow} \Bigg] \label{eq:15} \\
 \ket{\psi_{2}} & =  \Big( \frac{-a}{|a|} \Big) \Bigg[ b^{*}\Bigg( 1 + \imath \beta \Delta + \Delta^{2} \big[ \epsilon\big(\frac{2 |a|^{2} |b|^{2} + |b|^{2}}{2}\big) - \eta \big] \Bigg) \ket{\uparrow} - a^{*}  \Bigg(1 + \Delta^{2} \epsilon\Big(\frac{2|a|^{2} |b|^{2} -|b|^{2}}{2} \Big) \Bigg) \ket{\downarrow} \Bigg] \label{eq:16}
\end{align}
\end{widetext}
with associated eigenvalues:
\begin{align}
p_{1} & = 1 - |a|^{2}|b|^{2}\epsilon \Delta^{2}  \label{eq:17}\\
p_{2} & = |a|^{2}|b|^{2}\epsilon \Delta^{2} \label{eq:18}
\end{align}
where the parameter $\epsilon$ is defined by:
\begin{equation} \label{eq:19}
\epsilon \equiv \eta + \eta^{*} - \beta^{2}.
\end{equation}
\noindent We note that $\epsilon$ can also be written as
\begin{equation}
        \epsilon = 
        \frac{\lambda^{2}}{\hbar^{2}} \Big[ \bra{\phi_{e}} (H_{e}^{\downarrow} - H_{e}^{\uparrow})^{2} \ket{\phi_{e}} - \big(\bra{\phi_{e}} (H_{e}^{\downarrow} - H_{e}^{\uparrow}) \ket{\phi_{e}} \big)^{2} \Big] 
        \label{eq:variance}
\end{equation}
\noindent (using Eqns.~\ref{eq:13} and~\ref{eq:14}).  Equation~\ref{eq:variance} shows that $\epsilon$ is just the variance of $\frac{\lambda}{\hbar}(H_{e}^{\downarrow} - H_{e}^{\uparrow})$, so $\epsilon \geq 0$ by definition. Furthermore, $\epsilon = 0$ is a degenerate case where einselection does not occur---you can see, for example, that when $\epsilon = 0$ the second eigenvalue in Eqn.~\ref{eq:18} exactly disappears and the only state remaining with any probability is the initial state.  Thus, for cases of interest here $\epsilon>0$. Also note that $ \braket{\psi_1 | \psi_2} = \braket{\psi_2 | \psi_1} = 0 + O(\Delta^{3})$ and $ \braket{\psi_1 | \psi_1} = \braket{\psi_2 | \psi_2} = 1 + O(\Delta^{3})$, as you would expect from an $O(\Delta^{2})$ calculation.

Inspecting Eqns.~\ref{eq:15} and \ref{eq:16}, the zeroth order terms identify $\ket{\psi_{1}}$ with the original state of the system ($\ket{\psi_s(0)}$ from Eqn.~\ref{eq:6} apart from an irrelevant overall phase) and $\ket{\psi_{2}}$ as the orthogonal state.  In a two dimensional Hilbert space, there is (up to an overall phase) only one orthogonal state to $\ket{\psi_{1}}$.  From the way $a$ and $b$ alternate locations in the expressions for $\ket{\psi_{1}}$ vs $\ket{\psi_{2}}$, one can see that the two states have the ``mirror image" feature which led us to call the second eigenstate a``copycat" state in Fig.~\ref{fig:CopyCat}.  Thus, we see that at very early times the copycat profile is achieved automatically in this simple illustration.  

We noted in the introduction that the initial time evolution of the copycat state appeared to be ``slow'' in our numerical simulations.  This is also apparent in our analytic solutions---they show the time dependence of the eigenstates and their associated probabilities to be quadratic to leading order modulo a linear complex phase (which we have by convention placed in the coefficients of $\ket{\uparrow}$).  The quadratic early time dependence of the eigen\emph{values} has been anticipated before by calculations in \cite{KueblerZeh, JoosZeh}---where their ``rate of deseparation" is analogous to the quantity $|a|^{2}|b|^{2} \epsilon$---but the eigenstate solutions and their copycat nature is a new feature of our analysis.




\section{\label{sec:Implications} Further perturbative analysis }

\subsection{\label{sec:timeEvolv} Continued evolution of eigenstates}
To investigate what happens to the system density matrix eigenstates after the copycat state appears, let us again consider Eqns.~\ref{eq:15} and \ref{eq:16}. At the onset of the copycat process, for small $t = \Delta$, the system has already begun to decohere, but einselection has hardly started. While the system and environment are clearly entangled, the system density matrix eigenstates are not yet described by the pointer states $\ket{\uparrow}$ and $\ket{\downarrow}$ of the Hamiltonian in Eqn.~\ref{eqn:Hrcl}.

We now explore the perturbative behavior as $\Delta$ increases. Here we focus on $|\psi|^{2}$ of each eigenstate, given by
\begin{align}
\braket{\psi_{1} | \psi_{1}} & =  |a|^{2} \Big[  1 + \epsilon \Delta^{2}( 2 |a|^{2} |b|^{2} - |b|^{2} ) \Big] \braket{\uparrow | \uparrow} \notag \\
{} & \quad \quad + |b|^{2} \Big[  1 + \epsilon \Delta^{2} (2 |a|^{2} |b|^{2} - |a|^{2} ) \Big]\braket{\downarrow | \downarrow} \label{eq:21}\\
 \braket{\psi_{2} | \psi_{2}} & =  |b|^{2} \Big[  1 + \epsilon \Delta^{2}( 2 |a|^{2} |b|^{2} - |a|^{2} ) \Big] \braket{\uparrow | \uparrow} \notag \\
 {} & \quad \quad + |a|^{2} \Big[  1 + \epsilon \Delta^{2} (2 |a|^{2} |b|^{2} - |b|^{2} ) \Big]\braket{\downarrow | \downarrow} \label{eq:22}
\end{align}
where we have made use of Eqn.~\ref{eq:19}.
 
For $\epsilon > 0$---which is true in all cases where our analysis holds (see the discussion below Eqns.~\ref{eq:19} and~\ref{eq:variance})---we can re-write Eqns.~\ref{eq:21} and \ref{eq:22} as:
\begin{align}
\braket{\psi_{1} | \psi_{1}} & =  |a|^{2} \Big[  1 +  C_{1}^{\uparrow}(\Delta) \Big] \braket{\uparrow | \uparrow} + |b|^{2} \Big[  1 +  C_{1}^{\downarrow}(\Delta) \Big]\braket{\downarrow | \downarrow} \label{eq:23}\\
 \braket{\psi_{2} | \psi_{2}} & =  |b|^{2} \Big[  1 +  C_{2}^{\uparrow}(\Delta) \Big] \braket{\uparrow | \uparrow} + |a|^{2} \Big[  1 + C_{2}^{\downarrow}(\Delta) \Big]\braket{\downarrow | \downarrow} \label{eq:24}
\end{align}

\vspace{-0.3cm}

\noindent and construct the following chart for the sign of the time dependent coefficients as time increases:
\begin{center}
 \begin{tabular}{|c | c c c c|} 
 \hline
 Original State & $C_{1}^{\uparrow} (\Delta)$ & $C_{1}^{\downarrow} (\Delta)$ & $C_{2}^{\uparrow} (\Delta)$ & $C_{2}^{\downarrow} (\Delta)$ \\ [1ex] 
 \hline\hline
 $|a|^{2} > |b|^{2}$ & + & - & - & + \\ 
 \hline
  $|a|^{2} < |b|^{2}$ & - & + & + & - \\
 \hline
 $ |a|^{2} = |b|^{2} $ & 0 & 0 & 0 & 0 \\ [1ex] 
 \hline
\end{tabular}
\end{center}
Comparing Eqns.~\ref{eq:21} and \ref{eq:22} with Eqns.~\ref{eq:23}, \ref{eq:24}, and the chart illustrates that the subsequent evolution of the system eigenstates towards einselection is determined by the hierarchy of $|a|^{2}$ and $|b|^{2}$, the initial system probabilities---and that interactions with the environment control how fast this evolution occurs through the parameter $\epsilon$.

For example, suppose the initial state of the system is such that $|a|^{2} > |b|^{2}$ at $t=0$.  As $t = \Delta$ grows we see that the probability to observe $\ket{\psi_{1}}$ in the $\ket{\uparrow}$ state increases, since $C_{1}^{\uparrow} (\Delta)$ is increasing over time, and that the probability of observing $\ket{\psi_{1}}$ in the $\ket{\downarrow}$ state is decreasing by the same token.  The exact opposite trends occur in $\ket{\psi_{2}}$, the orthogonal state.  So long as $\braket{\psi_{1} | \psi_{1}} = \braket{\psi_{2} | \psi_{2}} = 1 + O(\Delta^{3})$ is preserved (namely that the perturbation expansion remains valid), the full system will exhibit these trends.  The system approaches complete einselection once $1 + C_{1}^{\downarrow} (\Delta) \approx  1 + C_{2}^{\uparrow} (\Delta)\approx 0$. An exactly analogous explanation occurs for the case of $|a|^{2} < |b|^{2}$. 

For the case of $|a|^{2} = |b|^{2}$, all time dependent coefficients vanish for a properly normalized state to leading order $O(\Delta^{2})$. One could interpret this as evidence for a static system---that after the onset of the copycat process no further evolution of the eigenstates occurs.  However, in the exactly degenerate limit $|a|^{2} = |b|^{2}$ all states are equally ``good" eigenstates of the system density matrix, so einselection into a specific basis of pointer states has no meaning in this limit.\footnote{In cases with very small deviations away from complete degeneracy, small irregularities (due, for example, to the finite size of the environment) can disrupt any tendency toward einselection.  We have seen this phenomenon in our numerical work, where for sufficiently degenerate cases finite size effects introduced large random fluctuations which dominated over the einselection process.} 

\subsection{\label{sec:decohTime} Decoherence Time}
Full einselection will take place on the timescale set by the decoherence processes. 
A system that has fully einselected will have the off-diagonal elements of its density matrix close to zero when $\rho_s$ is expressed in the pointer state basis~\cite{Zurek:2003zz}. Although this stage is only reached outside of the range of our perturbative calculations, we can still estimate the decoherence time by solving for the value of $\Delta$ where the off-diagonal elements are zero for our perturbative calculations. Applying this to Eqn.~\ref{eq:12} gives:
\begin{align}
 & \rho_{\ket{\uparrow} \bra{\downarrow}} =  a b^{*} \Big( 1 + \imath \beta \Delta - \eta \Delta^{2} \Big) = 0 \notag \\
& \rho_{\ket{\downarrow} \bra{\uparrow}} =  b a^{*} \Big( 1 - \imath \beta \Delta - \eta^{*} \Delta^{2} \Big)   = 0 \label{eq:25}
\end{align}
which can be rewritten as
\begin{align}
   & a b^{*} + b a^{*} + \imath \beta \Delta (a b^{*} - b a^{*}) - \Delta^{2}(a b^{*} \eta + b a^{*} \eta^{*}) = 0  \label{eq:26}\\
   & a b^{*} - b a^{*} + \imath \beta \Delta (a b^{*} + b a^{*}) - \Delta^{2}(a b^{*} \eta - b a^{*} \eta^{*}) = 0 \label{eq:27}.
\end{align}
Solving Eqns.~\ref{eq:26} and \ref{eq:27} together 
and simplifying yields the following perturbative estimate for the decoherence time:
\begin{equation} \label{eq:28}
    {\Delta_{d}} = \sqrt{\frac{2}{\eta + \eta^{*} }}
\end{equation}
where $\eta$ is given by Eqn.~\ref{eq:14}. Note that this result is independent of $a$ and $b$ (the initial state of the system).  We have compared this expression with our numerical work and found it gives reasonable estimates of the decoherence time. 
 
\subsection{\label{sec:spin} Spin Observables}
Here we consider the behavior of the Pauli spin operators in our RCL solutions. This will allow contact to be made with various experimental contexts such as NMR and quantum computing \cite{CPZ1,CPZ2,Schlosshauer}.

Our basis states for our system density matrix, $\{ \ket{\uparrow}, \ket{\downarrow} \}$, can be identified with the $S_z$ eigenbasis for spin-$
\frac{1}{2}$, so we can compute the expectation values for the spin operators $S_{x}$, $S_{y}$ and $S_{z}$ by 
\begin{equation} \label{eq:29}
    \braket{S_{i}} = Tr(\rho_{s}S_{i}) = \frac{\hbar}{2}Tr(\rho_{s} \sigma_{i})
\end{equation}
where the $\sigma_i$ are the usual Pauli matrices
\begin{equation} \label{eq:30}
 \sigma_x  = \begin{bmatrix} 0 & 1\\
   1 & 0
   \end{bmatrix} \quad
    \sigma_y  = \begin{bmatrix} 0 & - \imath \\
    \imath & 0
   \end{bmatrix} \quad
    \sigma_z  = \begin{bmatrix} 1 & 0\\
   0 & -1
   \end{bmatrix}.
\end{equation}
This gives
\begin{align}
    \braket{S_z} & = \frac{\hbar}{2}(|a|^{2} - |b|^{2}) \label{eq:31}\\
    \braket{S_x} & = \frac{\hbar}{2} \Big(2 Re[ab^{*}] - 2 \beta \Delta Im[ab^{*}] - 2\Delta^{2} Re[ab^{*}\eta]\Big) \label{eq:32}\\
    \braket{S_y} & = \frac{\hbar}{2} \Big(-2 Im[ab^{*}] - 2 \beta \Delta Re[ab^{*}] + 2\Delta^{2} Im[ab^{*}\eta]\Big). \label{eq:33}
\end{align}

Note that the system will have fully decohered/completed einselection when $\braket{S_x} = \braket{S_y} = 0$.  For our perturbative expressions, this condition is the same as that imposed by Eqns.~\ref{eq:26} and \ref{eq:27}.

\section{\label{sec:Beyond} Beyond Perturbation Theory}

An intriguing part of the copycat process is that its general features are agnostic about the spectrum of the environment.  The Hamiltonian in Eqn.~\ref{eqn:Hrcl} used to derive our results thus far makes no assumptions about the pieces that operate on the state of the environment, $H_{e}^{\uparrow}$ and $H_{e}^{\downarrow}$, except that they are time independent. This gives our results a flavor of generality often missing from canonical toy models in the literature---reviewed in \cite{Schlosshauer,BreuerPetruccione} and others---which typically make specific assumptions of ``ohmic'' environments and the like in order to arrive at concrete mathematical expressions.

However, if we do further specify $H_{e}^{\uparrow}$ and $H_{e}^{\downarrow}$ we can compute a non-perturbative version of the density matrix in Eqn.~\ref{eq:12}, closely following an approach by Zurek for a similar model \cite{Zurek1982}. This has two benefits.  First, it is possible to re-derive a form of the copycat results as leading order terms in the time series expansion of the non-perturbative solutions---as should be the case.  Second, a non-perturbative approach allows us to interpret the full time range of our numerical results discussed in \cite{ACLintro} from an analytic perspective.   

\subsection{\label{sec:AltDef} An Alternate Derivation}
To derive a non-perturbative version of Eqn.~\ref{eq:12}, we begin by specifying the following:
 \begin{align}
 H^{\uparrow}_{e} & = \sum^{N}_{i} \hbar \omega_{i} \ket{\omega_i} \bra{\omega_i} \label{eq:34}\\
 H^{\downarrow}_{e} & = \sum^{N}_{i} \hbar f^{\uparrow \downarrow}\omega_{i} \ket{\omega_i} \bra{\omega_i} \label{eq:35}
 \end{align}
 so that
 \begin{equation} \label{eq:36}
 \omega^{\downarrow}_{i} = f^{\uparrow \downarrow} \omega^{\uparrow}_{i}
 \end{equation}
i.e. $H_{e}^{\uparrow}$ and $H_{e}^{\downarrow}$ are almost identical, except for a tuneable real dimensionless constant $f^{\uparrow \downarrow}$. (The RCL model discussed in~\cite{ACLintro} has this form with $f^{\uparrow \downarrow}=-1$.) This simplifying assumption about the relationship between $H_{e}^{\uparrow}$ and $H_{e}^{\downarrow}$ enables the analysis which follows.

We can express the state of the environment in the energy eigenbasis of  $H_{e}^{\uparrow}$ and $H_{e}^{\downarrow}$, so that
\begin{equation} \label{eq:37}
\ket{\phi_{e}} = \sum^{N}_{i} \alpha_{i} \ket{\omega_i}
\end{equation}
\noindent with the normalization condition:
\begin{equation} \label{eq:38}
\sum^{N}_{i} |\alpha^{\uparrow}_{i}|^{2} = 1.
 \end{equation}

Both the ACL and RCL models operate within a finite dimensional Hilbert space, so we have made that explicit in our forms for $H_{e}^{\uparrow}$ and $H_{e}^{\downarrow}$.  The exact frequency spectrum of the  $\omega_{i}$'s is still arbitrary. However, when comparisons with our numerical work are made in the next section, we will take their distribution to be random and centered around zero---to coincide with the random nature of $H_{e}^{\uparrow}$ and $H_{e}^{\downarrow}$ in the ACL model \cite{ACLintro}.
 
Given the definitions in Eqns.~\ref{eq:34} - \ref{eq:38}, we can re-express the original state of the system and environment as:
\begin{equation} \label{eq:39}
    \ket{\psi_{w}(0)} = (a\ket{\uparrow} + b\ket{\downarrow}) \otimes \sum_{i}^{N} \alpha_{i} \ket{\omega_{i}}
\end{equation}
\noindent and the RCL Hamiltonian originally given in Eqn.~\ref{eqn:Hrcl} becomes:
\begin{align}
    H_{w} = & (\lambda \ket{\uparrow} \bra{\uparrow} \otimes \sum_{i}^{N} \hbar \omega_{i}\ket{\omega_{i}} \bra{\omega_{i}}) \notag \\
    & {} + (\lambda \ket{\downarrow} \bra{\downarrow} \otimes \sum_{i}^{N} \hbar \omega_{i} f^{\uparrow \downarrow} \ket{\omega_{i}} \bra{\omega_{i}})  \label{eq:40}
\end{align}

Since we have made the eigenvalues of our Hamiltonian explicit from the start, we may write down the full time evolved state as:
\begin{align}
    \ket{\psi_{w}(t)} = & (a \sum_{i}^{N} \alpha_{i} e^{- \imath \lambda \omega_{i} t} \ket{\uparrow} \otimes \ket{\omega_{i}}) \notag \\
   & {}\quad  + (b \sum_{i}^{N} \alpha_{i} e^{- \imath \lambda f^{\uparrow \downarrow} \omega_{i} t} \ket{\downarrow} \otimes \ket{\omega_{i}}) \label{eq:41}
\end{align}
\noindent Tracing over the environment then gives the following system density matrix:
\begin{align}
    \rho_{s}(t) = & |a|^{2} \ket{\uparrow} \bra{\uparrow} + a b^{*} z(t)  \ket{\uparrow} \bra{\downarrow} \notag \\
    & {} + b a^{*} z^{*}(t)  \ket{\downarrow} \bra{\uparrow} + |b|^{2} \ket{\downarrow} \bra{\downarrow} \label{eq:42}
\end{align}
\noindent where the quantity $z(t)$ has been called the correlation amplitude or decoherence factor \cite{Zurek1982, Zurek:2003zz,CPZ1,CPZ2} for similar toy models and is given in our notation by:
\begin{equation} \label{eq:43}
    z(t) = \sum_{i}^{N} |\alpha_{i}|^{2} e^{- \imath t \lambda \omega_{i} (1 - f^{\uparrow \downarrow})}
\end{equation}
Note that $z(t)$ is a sum of complex exponentials that directly depends on the difference in eigenvalues of our two environmental Hamiltonians. As Zurek originally discussed in  \cite{Zurek1982}, the quantity $|\alpha_{i}|^{2}$ describes the probability of finding the environment in the different eigenstates of the interaction Hamiltonian, and it is possible to show that the average value of $z(t)$ will approach zero for sufficiently long times, effectively damping out the off-diagonal system density matrix elements.

Together Eqns.~\ref{eq:42} and \ref{eq:43} are the non-perturbative version of the system density matrix in Eqn.~\ref{eq:12}, for the specific realization of environment parameters given by Eqns.~\ref{eq:34} - \ref{eq:38}. To show how these results connect with the copycat process solutions, first write $z(t)$ in terms of trigonometric functions.
\begin{equation} \label{eq:44}
     z(t) = \sum_{i}^{N} |\alpha_{i}|^{2} \big[ \cos(\lambda \omega_{i} (1 - f^{\uparrow \downarrow}) t) - \imath \sin(\lambda \omega_{i} (1 - f^{\uparrow \downarrow}) t) \big]
\end{equation}
\noindent Then we take the limit $t \rightarrow \Delta$ by keeping only the first non-trivial term in each trigonometic function's series expansion. We still keep the sum over the states of the environment, all we are doing is an early time expansion. This yields:
\begin{align}
    z(\Delta) = & 1 + \imath \sum_{i}^{N} |\alpha_{i}|^{2} \lambda \omega_{i} (f^{\uparrow \downarrow} - 1) \Delta \notag \\
    & {} \qquad - \sum_{i}^{N} |\alpha_{i}|^{2} \omega_{i}^{2} \frac{ \lambda^{2} (f^{\uparrow \downarrow} -1)^{2} }{2} \Delta^{2} \notag \\
    = & 1 + \imath \beta \Delta - \eta \Delta^{2} \label{eq:45}
\end{align}
\noindent which is exactly the time dependent off-diagonal element in Eqn.~\ref{eq:12}, given the definitions in Eqns.~\ref{eq:34} - \ref{eq:38}.  One can verify the equivalence between the two lines of Eqn.~\ref{eq:45} by starting with the definitions of $\beta$ and $\eta$ given by Eqns.~\ref{eq:13} and \ref{eq:14}, and then substituting in the specific forms of $H_{e}^{\uparrow}$, $H_{e}^{\downarrow}$, and $\ket{\phi_{e}}$ given in this section---the result will be the same as Eqn.~\ref{eq:45}. An analogous expression for $z^{*}(\Delta)$ holds, which enables us to re-express Eqn.~\ref{eq:12} as
\begin{align}
    \rho_{s}(\Delta) = & |a|^{2} \ket{\uparrow} \bra{\uparrow} + a b^{*} z(\Delta)  \ket{\uparrow} \bra{\downarrow} \notag \\
    & {} + b a^{*} z^{*}(\Delta)  \ket{\downarrow} \bra{\uparrow} + |b|^{2} \ket{\downarrow} \bra{\downarrow} \label{eq:46}
\end{align}
From Eqns.~\ref{eq:45} and \ref{eq:46}, one can then go on to determine the eigenvalues, eigenvectors and decoherence time of the system density matrix.  The results will match the more general calculations in Sections III and IV, for the specific versions of $\eta$ and $\beta$ given in Eqn.~\ref{eq:45}.

To summarize, in this section we have derived a specific realization of the copycat results as leading order terms in the time series expansion of Eqns.~\ref{eq:42} and \ref{eq:43}. Note that if the relationship between $H_{e}^{\uparrow}$ and $H_{e}^{\downarrow}$ in Eqns.~\ref{eq:34} and \ref{eq:35} was more complicated---if they did not share the same energy eigenbasis or if the relationship between eigenvalues was non-linear, for example---then a derivation of the early time density matrix from a non-perturbative approach might not proceed as smoothly as we just described. However, the early time results of Sect.~\ref{sec:CCintro} will have a more general range of validity. 

\subsection{\label{sec:Numerical} Comparison with numerical results}
In this section we consider two quantities that depend strongly on the off-diagonal elements of the system density matrix---the linear entropy and $\braket{S_{x}}$---and 
compare numerical non-perturbative results to semi-analytic early-time expressions. 

The linear entropy for a density matrix is defined as:
\begin{equation}\label{eq:47}
    S_{l}(\rho) = 1 - Tr[\rho^{2}]
\end{equation}
which is bounded according to $0 \leq S_{l} \leq 1$ \cite{BreuerPetruccione}.  For the system density matrix given in our perturbative analysis (Eqn.~\ref{eq:12}), this yields:
\begin{equation} \label{eq:48}
    S_{l,P} = 2 |a|^{2}|b|^{2}\epsilon \Delta^{2} + O(\Delta^{3})
\end{equation}
with $\epsilon$ given by:
\begin{equation} \label{eq:49}
\epsilon  =  \lambda ^{2} (f^{\uparrow \downarrow} - 1)^{2}\Bigg[\sum^{N}_{i} \omega_{i}^{2} |\alpha_{i}|^{2} - \Bigg(\sum^{N}_{i}  \omega_{i} |\alpha_{i}|^{2} \Bigg)^{2} \Bigg]
\end{equation}
\noindent assuming Eqns.~\ref{eq:34} - \ref{eq:38} from the previous section.

For the non-perturbative case, given Eqns.~\ref{eq:42} and \ref{eq:43}, one obtains:
\begin{equation} \label{eq:50}
    S_{l,NP} = 1 - |a|^{4} - |b|^{4} - 2 |a|^{2}|b|^{2} |z(t)|^{2}
\end{equation}
with
\begin{equation} \label{eq:51}
    |z(t)|^{2} = \sum_{i,j}^{N} |\alpha_{i}|^{2} |\alpha_{j}|^{2} e^{- \imath \lambda (1 - f^{\uparrow \downarrow}) (\omega_{i} - \omega_{j}) t } 
\end{equation}

\begin{figure}[h!]
\centering
\includegraphics[width=3.5in]{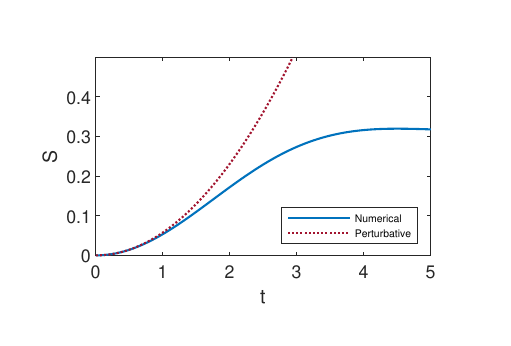}
\caption{Linear entropy curves. Solid: Non-perturbative (corresponding to numerically evaluating Eqns.~\ref{eq:50} and~\ref{eq:51}). Dotted: Perturbative expression from Eqn.~\ref{eq:48}, using $\epsilon$ (given by Eqn.~\ref{eq:49}) drawn from the same numerical calculation shown.}
\label{fig:S}
\end{figure}

Figure~\ref{fig:S} shows the perturbative and non-perturbative linear entropies as functions of time. To numerically generate the non-perturbative solid curve, our simulations effectively evaluate the summation in Eqn.~\ref{eq:51} followed by Eqn.~\ref{eq:50} at each time-step and plot the result.  For the dotted curve corresponding to Eqn.~\ref{eq:48}, the summation in Eqn.~\ref{eq:49} is evaluated once numerically and then the expression in Eqn.~\ref{eq:48} is plotted for the same time-steps as those used for Eqn.~\ref{eq:50}. For both curves: $a = 1/\sqrt{5}$, $b = 2/\sqrt{5}$, $f^{\uparrow \downarrow}=-1$, and the distribution of environmental frequencies,  $\omega_{i}$, is taken to be random and centered around zero---to coincide with the random nature of $H_{e}^{\uparrow}$ and $H_{e}^{\downarrow}$ in the ACL model \cite{ACLintro}.

Next, consider $\braket{S_{x}}$. For the early-time regime we simply have Eqn.~\ref{eq:32}, reprinted here:
\begin{equation} \label{eq:52}
   \braket{S_{x}}_P  = \frac{\hbar}{2} \Big(2 Re[ab^{*}] - 2 \beta \Delta Im[ab^{*}] - 2\Delta^{2} Re[ab^{*}\eta]\Big)
\end{equation}
\noindent with the quantities $\eta$ and $\beta$ as defined in Eqn.~\ref{eq:45}.  For the non-perturbative case we obtain:
\begin{equation} \label{eq:53}
    \braket{S_{x}}_{NP} = \frac{\hbar}{2} (2 Re[ab^{*}z(t)])
\end{equation}
\noindent with $z(t)$ given by Eqn.~\ref{eq:43}.

\begin{figure}[h!]
\centering
\includegraphics[width=3.4in]{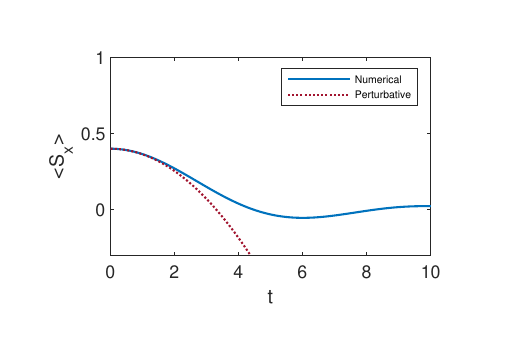}
\caption{$\braket{S_x}$, giving the real part of the off-diagonal element of $\rho_{s}$. The solid curve is non-perturbative, corresponding to numerically evaluating Eqn~\ref{eq:53}. The dotted curve corresponds to the early time analytic expression in Eqn.~\ref{eq:52}, with the the quantities defined in Eqn~\ref{eq:45} drawn from the numerical calculation. We take $\hbar = 1$.}
\label{fig:Sx}
\end{figure}

Figure~\ref{fig:Sx} shows the perturbative and non-perturbative results for $\braket{S_{x}}$ as a function of time.  As with the linear entropy, the non-perturbative curve was generated numerically in our simulations essentially by evaluating Eqn.~\ref{eq:53} for each time step, while the  early-time result is the analytical function in Eqn.~\ref{eq:52} with the summations for $\eta$ and $\beta$ in Eqn.~\ref{eq:45} evaluated numerically. The values for $a$, $b$, $f^{\uparrow \downarrow}$, and $\omega_{i}$ are the same as in Fig.~\ref{fig:S}, and we take $\hbar = 1$.

We note here that the RCL model is not highly efficient at completing the process of einselection, as evidenced by the small oscillation around zero of the numerical curve in Fig.~\ref{fig:Sx}.  We further discuss the interpretation of these oscillations in~\cite{ACLintro} and link them to phenomena seen in NMR experiments.  We also identify a modification to the RCL model which reduces these oscillations, thereby further illuminating their physical origins. 

Both Figs.~\ref{fig:S} and \ref{fig:Sx} demonstrate that the perturbative regime---characterized by the copycat density matrix in Eqn.~\ref{eq:12}---can be a sizable portion of the full time evolution of the system's linear entropy and $\braket{S_{x}}$. The duration may vary somewhat for different system-environment coupling strengths or environmental frequency spectra, but the overall presence of a significant period of quadratic time behavior is clear.

\section{\label{sec:conclude}Discussion and Conclusions}
In this paper we have identified and described unique early time behavior of a quantum system interacting with its environment---the copycat process. The copycat process is a new and potentially important addition to the narrative of decoherence and einselection. By considering the evolution of the system density matrix from an eigenstate perspective, we were able to recognize the early-time emergence of a distinct transiently stable ``copycat'' state, as illustrated in Fig.~\ref{fig:CopyCat}.  We have derived the same effect analytically in Sect.~\ref{sec:CCintro}, and then utilize the solutions and their implications to obtain new insights into how small quantum systems einselect in Sects.~\ref{sec:Implications} and~\ref{sec:Beyond}.  Furthermore, the comparison with our numerical work in section V.B demonstrates that one might expect key features of the copycat process to dominate for a significant portion of the full time to full einselection.

The generality of our results in Sects.~\ref{sec:CCintro} and~\ref{sec:Implications} is also noteworthy. As we briefly commented in Sect.~\ref{sec:Beyond}, an intriguing part of the copycat process is that it is agnostic about the spectrum of the environment.  The Hamiltonian in Eqn.~\ref{eqn:Hrcl} used to derive the copycat results makes no assumption of any of the standard environmental spectra---such as ``ohmic'' environments---typically employed in the literature to make analytical progress \cite{Schlosshauer,BreuerPetruccione}, and it also does not assume the random environment that we utilize for our numerical work in section V.B and \cite{ACLintro}.  This suggests that the onset of einselection could begin with the copycat process in a wide variety of cases.

While we acknowledge our analytical modelling of system-environment interactions is fairly simplified in the RCL model, our numerical work with the ACL model in  \cite{ACLintro} demonstrates that the copycat process persists even in the presence of strong self Hamiltonians of the system and environment. Furthermore, we expect the copycat process to be present in some form for larger and more complicated systems.  The orthogonal nature of the copycat eigenstates might be unsurprising for the two-state system results---the small Hilbert space greatly limits the possibilities---but we have also seen that the same copycat behavior holds for superpositions of two coherent states of an SHO. 

When the evolution of the global ($w$) system is unitary, the evolution of a subsystem density matrix is always deterministic.  Thus (except for the case of degenerate eigenvalues) the evolution the density matrix eigenstates is also deterministic.  The copycat process is an example of a form of this deterministic evolution which generically appears in two state systems, as well as some larger systems that are started in ``two cat'' states. 

To explore further, in Appendix~\ref{sec:qtrit} we extend our analytical perturbative analysis to the case where the system is a qu\emph{trit}.  There we see that many of the same copycat features appear, although the pattern of early quadratic behavior is partially broken by the possibility of linear evolution in the (2d) system subspace orthogonal to the original system state.  

\begin{figure}[h!]
\centering
\includegraphics[width=3.4in]{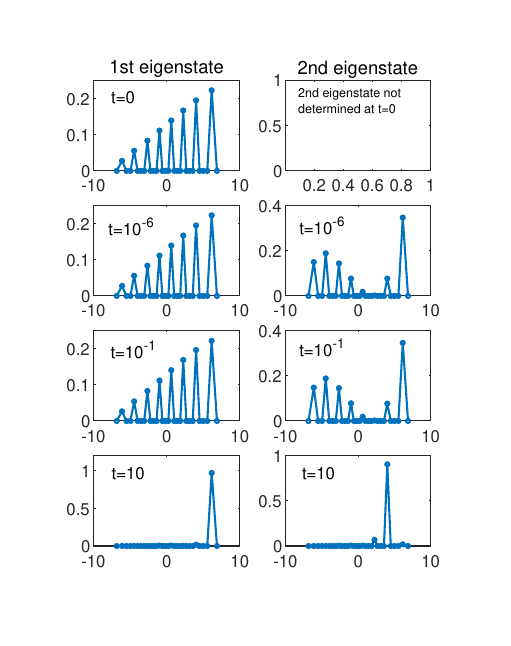}
\caption{The evolution of an initial eight-cat state, showing the top two eigenstates of $\rho_s$. The term ``copycat'' might not be a great description of the second eigenstate, but we have found that the quadratic transient stability is still present.  The process of einselection is essentially complete at the final time shown. We use the RCL model, with a $d=30$ qudit system.  The markers show the amplitude squared for each basis vector, and the lines are added for illustrative purposes.  }
\label{fig:Fig8}
\end{figure}

Exploring further still, Fig.~\ref{fig:Fig8} shows the evolution of a particular example with a higher dimensional system, set up in a form that might be thought of as ``eight Schr\"odinger cats.''  In that case we did observe overall quadratic behavior to leading order, resulting in the same transient stability.  One would be hard pressed to describe the states that appear in the right column as ``copycat states,'' but these are the states which emerge from the deterministic Schr\"odinger evolution in the $w$ space for the particular chosen initial state.   
These results give some sense how the equivalent process can look in a more complex situation. 

There are several connections between the results we present here and the existing literature. Early-time quadratic decay of the off-diagonal elements of the system density matrix for decohering systems has been mentioned to varying degrees in several places~\cite{Schlosshauer,BreuerPetruccione,Unruh:1994az,Palma:1997gs,Zurek1982,CPZ1,CPZ2,JoosZeh,KueblerZeh,Dobrovitskietal,BraunetAl}.  However, these references typically do not look explicitly at the eigenstates of $\rho_s$.
Other explorations of the early-time behavior of open quantum systems appear in discussions of the ``quantum Zeno paradox'' \cite{BreuerPetruccione,Joos1984,Peres:1980ux}, where the behavior of the system density matrix eigenstates is also typically not considered.  

Generally, we have found that features of the evolution that might be associated with ``decay,'' or the onset of entanglement, are quadratic to leading order.  Such features are controlled by the eigen\emph{values} of $\rho_s$, which are always quadratic (to leading order) in our results and in the literature we cite here. 
We have also quite generally found aspects of the evolution which are \emph{linear} at lowest order in time. These aspects describe evolution of the system in ways not associated with the onset of entanglement. For example, the linear piece $\propto\beta$ in Eqns.~\ref{eq:15} and~\ref{eq:16} describes the evolution of the relative phase between the coefficients of the pointer states, and in Appendix~\ref{sec:qtrit} we saw a real linear part to the evolution of the second and third eigenstates of $\rho_s$.

We also note that many approaches to studying decoherence and einselection---see reviews in \cite{Schlosshauer,BreuerPetruccione}, for example---utilize a master equation approach to analyze the time evolution of the system density matrix.  As discussed in \cite{ACLintro}, this master equation approach typically carries with it assumptions of Markovian evolution and the resulting exponential decay of off-diagonal system density matrix elements. Even within the master equation approach it is known that exponential decay is not always valid \cite{Anglin_1997,BreuerPetruccione,Schlosshauer}, however the exponential case remains the focus of much of the literature. There are some exceptions to this focus. 
Zurek and collaborators~\cite{Zurek1982,CPZ1,CPZ2} explicitly note the generally dominant early quadratic behaviors and point out that, in the context of the formalisms they develop, the exponential behavior is a very special case. And Peres~\cite{PERES198033} offers a general analysis of the diverse range of possible behaviors. Our approach in this paper and in our numerical work \cite{ACLintro} is agnostic of Markovian assumptions by simply solving the Schr\"odinger equation directly.  This has led to us observing more complicated non-Markovian time dependence in our system density matrix, including the copycat regime.

Looking forward, we are curious whether our calculations of the copycat process could provide a useful tool for studying decoherence and einselection in open quantum systems. Examining the time dependence of the system density matrix eigenstates and eigenvalues allows one to see the system smoothly transition from an initial quantum superposition to a classical mixture of pointer states, with the copycat process describing the first stages of this transition. 

\vspace{-0.2cm}

\section{\label{sec:thanks}Acknowledgements}

We are grateful to F. Anza, N. Curro, P. Coles, B. Nachtergaele, R. Singh, A. Sornborger, and Z. Wang for valuable conversations. This work was supported in part by the U.S. Department of Energy, Office of Science, Office of High Energy Physics QuantISED program under Contract No. KA2401032.

\appendix

\section{\label{sec:qtrit}The qutrit RCL}
Here we extend our perturbative treatment of the RCL model to the case where the single qubit system is replaced with a qu\emph{trit}.  This enables us, among other things, to study the evolution of three cat Schr\"odinger cat initial states.  

Our results show that the early time behavior of the eigenvalues is quadratic to leading order, as we have already shown analytically in the qubit case and have also observed numerically in much larger systems.  In many respects the behavior of the eigenstates also reflects what we saw for the qubit case.  However, we have identified circumstances where the leading behavior of the second and third eigenstates has a real linear contribution, in contrast to the qubit case where the linear piece just showed up in a relative phase.  

The derivation is nearly identical to the treatment in Section III, in that we begin with an initial state
\begin{equation} \label{eq:B1}
 \ket{\psi_{w} (0)} = (a\ket{1} + b\ket{0} + c\ket{-1})\ket{\phi_{e}}
 \end{equation}
 with $|a|^{2} + |b|^{2} + |c|^{2} = 1$ and Hamiltonian of the form\footnote{Note that the ``$-1$'' superscript here is an index, not an inverse operation.}
 \begin{equation} \label{eq:B2}
H_{w} = \lambda \big(\ket{1}\bra{1} H^{1}_{e} + \ket{0}\bra{0} H^{0}_{e} + \ket{-1}\bra{-1} H^{-1}_{e}\big).
\end{equation}
Following the same methods as Section III, we obtained the reduced density matrix
\begin{equation}\label{eq:B3}
\rho_{s} = \begin{bmatrix}  \rho_{\ket{1} \bra{1}} & \rho_{\ket{1} \bra{0}} &  \rho_{\ket{1} \bra{-1}}\\
\rho_{\ket{0} \bra{1}} & \rho_{\ket{0} \bra{0}} &  \rho_{\ket{0} \bra{-1}}\\
\rho_{\ket{-1} \bra{1}} & \rho_{\ket{-1} \bra{0}} &  \rho_{\ket{-1} \bra{-1}}
\end{bmatrix}
\end{equation}
with entries defined as: 
\begin{align}
    &\rho_{\ket{1} \bra{1}} = a a^{*} \notag \\
    &\rho_{\ket{1} \bra{0}} = a b^{*} \big(1 + \imath \Delta \beta_{10} - \Delta^{2} \eta_{10} + \imath \Delta^{3} \nu_{10} + \Delta^{4} \kappa_{10} \big) \notag\\
    &\rho_{\ket{1} \bra{-1}}  = a c^{*} \big(1 + \imath \Delta \beta_{1-1} - \Delta^{2} \eta_{1-1} + \imath \Delta^{3} \nu_{1-1} + \Delta^{4} \kappa_{1-1} \big) \notag \\
    &\rho_{\ket{0} \bra{1}}  =  b a^{*} \big(1 - \imath \Delta \beta_{10} - \Delta^{2} \eta^{*}_{10} - \imath \Delta^{3} \nu^{*}_{10} + \Delta^{4} \kappa^{*}_{10}\big) \notag \\
    &\rho_{\ket{0} \bra{0}} = b b^{*} \notag \\
    &\rho_{\ket{0} \bra{-1}}  = b c^{*} \big(1 + \imath \Delta \beta_{0-1} - \Delta^{2} \eta_{0-1} + \imath \Delta^{3} \nu_{0-1} + \Delta^{4} \kappa_{0-1} \big) \notag\\
    &\rho_{\ket{-1} \bra{1}}  = c a^{*} \big(1 - \imath \Delta \beta_{1-1} - \Delta^{2} \eta^{*}_{1-1} - \imath \Delta^{3} \nu^{*}_{1-1} + \Delta^{4} \kappa^{*}_{1-1}\big) \notag \\
    &\rho_{\ket{-1} \bra{0}}  = c b^{*} \big(1 - \imath \Delta \beta_{0-1} - \Delta^{2} \eta^{*}_{0-1} - \imath \Delta^{3} \nu^{*}_{0-1} + \Delta^{4} \kappa^{*}_{0-1}\big) \notag \\
    &\rho_{\ket{-1} \bra{-1}}  = c c^{*}. \label{eq:B4}
\end{align}
It ends up being necessary to calculate the reduced density matrix to $O(\Delta^{4})$, in order to not lose information when calculating the eigenvalues and eigenstates to $O(\Delta^{2})$. The $\beta$ and $\eta$ parameters are defined analogously to Eqns.~\ref{eq:13} and \ref{eq:14}, i.e.
\begin{align}    
\beta_{ij} & =  \frac{\lambda}{\hbar}\big(\bra{\phi_{e}}  H^{j}_{e}\ket{\phi_{e}} - \bra{\phi_{e}} H^{i}_{e} \ket{\phi_{e}} \big) \label{eq:B5}\\
\eta_{ij} &  =  \frac{\lambda^{2}}{\hbar^{2}} \Big( \frac{\bra{\phi_{e}}H^{i}_{e}H^{i}_{e}\ket{\phi_{e}} + \bra{\phi_{e}}H^{j}_{e}H^{j}_{e}\ket{\phi_{e}}}{2} \notag \\
& \qquad \quad  - \bra{\phi_{e}} H^{j}_{e}H^{i}_{e} \ket{\phi_{e}} \Big) \label{eq:B6}
\end{align}
with 
\begin{equation} \label{eq:B7}
\epsilon_{ij} = \eta_{ij} + \eta^{*}_{ij} - \beta_{ij}^{2}
\end{equation}
\noindent and the additional third and forth order parameters, $\nu_{ij}$ and $\kappa_{ij}$, are defined according to:
\begin{align}
   \nu_{ij} &= \frac{\lambda^{3}}{\hbar^{3}}  \Big( \frac{\bra{\phi_{e}}(H^{i}_{e})^{3}\ket{\phi_{e}} - \bra{\phi_{e}}(H^{j}_{e})^{3}\ket{\phi_{e}}}{6} \notag \\
   & \quad +  \frac{\bra{\phi_{e}}(H^{j}_{e})^{2}H^{i}_{e}\ket{\phi_{e}} - \bra{\phi_{e}}H^{j}_{e}(H^{i}_{e})^{2}\ket{\phi_{e}}}{2} \Big) \label{eq:B8} \\
   \kappa_{ij} &= \frac{\lambda^{4}}{\hbar^{4}}  \Big( \frac{\bra{\phi_{e}}(H^{i}_{e})^{4}\ket{\phi_{e}} + \bra{\phi_{e}}(H^{j}_{e})^{4}\ket{\phi_{e}}}{24} \notag \\
   &  \quad -  \frac{\bra{\phi_{e}}(H^{j}_{e})^{3}H^{i}_{e}\ket{\phi_{e}} + \bra{\phi_{e}}H^{j}_{e}(H^{i}_{e})^{3}\ket{\phi_{e}}}{6}  \notag \\
   & \quad \qquad + \frac{\bra{\phi_{e}}(H^{j}_{e})^{2}(H^{i}_{e})^{2}\ket{\phi_{e}}}{4} \Big). \label{eq:B9} 
\end{align}
Because this 3 x 3 system reduced density matrix is Hermitian, general analytical solutions for the eigenvalues and eigenstates exist \cite{hal2017,Kopp:2006wp}. Using these exact solutions as a starting point, we then performed another sequence of series expansions for the small parameter $t =\Delta$ to obtain analytic solutions.  For the sequence of series expansions we kept terms up to $O(\Delta^{4})$, only truncating the results to $O(\Delta^{2})$ at the end.  As mentioned earlier, this is essential to not lose information when calculating the eigenvalues and eigenstates to $O(\Delta^{2})$---for example, one needs to keep up to $O(\Delta^{4})$ to navigate the series expansion of the ratio involving a square root in Eqn.~9 of \cite{hal2017} correctly.  The eigenvalues to lowest order in time are given by:
\begin{align}
    p_1 & = 1 - \Delta^{2} \Big(|a|^{2}|b|^{2}\epsilon_{10} + |a|^{2}|c|^{2}\epsilon_{1-1} + |b|^{2}|c|^{2}\epsilon_{0-1} \Big) \notag \\
    {} & = 1 - \Delta^{2} \lambda_{1} \label{eq:B10} \\
    p_2 & = \frac{\Delta^{2}}{2} \Big( \lambda_{1} +  \sqrt{\Lambda} \Big) \notag \\
    {} & = \Delta^{2} \lambda_{2} \label{eq:B11}\\
    p_3 & = \frac{\Delta^{2}}{2} \Big( \lambda_{1} - \sqrt{\Lambda} \Big) \notag   \\
    {} & = \Delta^{2} \lambda_{3} \label{eq:B12}
\end{align}
with
\begin{align}
    \Lambda & = \lambda_{1}^{2} + 4 |a|^{2}|b|^{2}|c|^{2} \Big( |\eta_{0-1}|^{2} +|\eta_{10}|^{2} + |\eta_{1-1}|^{2} \notag \\
    & \qquad + \beta_{10}\beta_{1-1}\text{Re}[\eta_{0-1}] + \beta_{0-1}\beta_{1-1}\text{Re}[\eta_{10}] \notag \\
    & \qquad - \beta_{0-1}\beta_{10}\text{Re}[\eta_{1-1}] - \text{Re}[\eta_{0-1}\eta_{10}] \notag \\
    & \qquad - \text{Re}[\eta_{1-1}\eta_{0-1}^{*}] - \text{Re}[\eta_{1-1}\eta_{10}^{*}] \Big). \label{eq:B13}
\end{align}
To obtain the above it is necessary to recognize that
\begin{equation} \label{eq:B14}
(\beta_{0-1} + \beta_{10} - \beta_{1-1}) = 0
\end{equation}
\noindent generically, simply following from the definition in Eqn.~\ref{eq:B5}. This sets the leading order time dependence of the eigenvalues to be quadratic, as with the two-state results earlier in this paper. Note the $\pm \sqrt{\Lambda}$ part of Eqns.~\ref{eq:B11} and~\ref{eq:B12} is what saves $p_2$ and $p_3$ from being degenerate at $O(\Delta^{2})$.
In the limit that any of the initial state coefficients $a$, $b$, or $c$ are sent to zero, we exactly recover the two state eigenvalues given by Eqns.~\ref{eq:17} and \ref{eq:18} from Eqns.~\ref{eq:B10} - \ref{eq:B13}.  

The normalized eigenstate results have the general form:
\begin{align}
   & \ket{\psi_{1}} = \chi_{1} \ket{1} + \gamma_{1} \ket{0} + \zeta_{1} \ket{-1} \label{eq:B15} \\
  & \ket{\psi_{2,3}} = \chi_{2,3} \ket{1} + \gamma_{2,3} \ket{0} + \zeta_{2,3} \ket{-1} \label{eq:B16}
\end{align}
where for the top eigenstate $\chi$, $\gamma$, and $\zeta$ are defined by:
\begin{align}
    \chi_{1} & = \frac{x_{0}}{N_{0}} \Big[ 1 + \imath \, \frac{x_{1}}{x_{0}} \Delta + \big(\frac{x_{2}}{x_{0}} - \frac{1}{2}\frac{N_{2}}{(N_{0})^{2}} \big) \Delta^{2} \Big] \label{eq:B17}\\
    \gamma_{1} & = \frac{y_{0}}{N_{0}} \Big[ 1 + \imath \, \frac{y_{1}}{y_{0}} \Delta + \big(\frac{y_{2}}{y_{0}} - \frac{1}{2}\frac{N_{2}}{(N_{0})^{2}} \big) \Delta^{2} \Big] \label{eq:B18}\\
    \zeta_{1} & = \frac{1}{N_{0}} \Big[ 1 - \frac{1}{2} \frac{N_{2}}{(N_{0})^{2}} \Delta^{2} \Big] \label{eq:B19}
\end{align}
given
\begin{align}
    N_{0} & = \sqrt{1 + |x_{0}|^{2} + |y_{0}|^{2}} \label{eq:B20} \\
    N_{2} & = |x_{1}|^{2} + |y_{1}|^{2} + 2 \text{Re}[x_{0} x_{2}^{*}] + 2 \text{Re}[y_{0} y_{2}^{*}] \label{eq:B21}
\end{align}
with
\begin{align}
    y_{0} & = \frac{b}{c} \label{eq:B22} \\
    y_{1} & = \frac{b}{c} \big(\beta_{1-1}  -\beta_{10} \big)  \label{eq:B23} \\
    y_{2} & = \frac{b}{c} \Big[ (\beta_{10} - \beta_{1-1})\beta_{1-1} +\eta_{1-1}^{*} - \eta_{10}^{*} \notag \\
    & \quad {} + |c|^{2}(\delta_{1} - \epsilon_{0-1}) + |b|^{2}\delta_{1} \Big] \label{eq:B24}
\end{align}
and
\begin{align}
  x_{0} & = \frac{a}{c} \label{eq:B25}  \\
  x_{1} & = \frac{a}{c} \beta_{1-1}  \label{eq:B26} \\
  x_{2} & = \frac{a}{c} \frac{1}{|a|^{2}} \Big[ (1 - |c|^{2}) (\epsilon_{1-1} - \eta_{1-1}) - \lambda_{1} - c b^{*} y_{2} \notag \\
  & \quad \quad \quad + |b|^{2} \big[ \beta_{0-1}^{2} - 2 \beta_{0-1}\beta_{1-1}  - \epsilon_{1-1} + \eta_{1-1} + \eta_{0-1}^{*} \big] \Big] \label{eq:B27}
\end{align}
where $\lambda_{1}$ is defined in Eqn.~\ref{eq:B10}, $y_2$ in Eqn.~\ref{eq:B24}, and $\delta_{1}$ is shorthand for:
\begin{equation} \label{eq:B28}
    \delta_{1} = \beta_{0-1}\beta_{10} + \eta_{0-1}^{*} + \eta_{10}^{*} - \eta_{1-1}^{*}.
\end{equation}
\noindent As with the two-state solutions, it is purely a matter of our chosen convention (chosen for convenience) that the linear complex phase is present in $\ket{1}$ and $\ket{0}$ but not $\ket{-1}$ in Eqns.~\ref{eq:B17} - \ref{eq:B19}.

Additional complexity is present for $\ket{\psi_{2}}$ and $\ket{\psi_{3}}$.  For these eigenstates, $\chi_{2,3}$, $\gamma_{2,3}$, and $\zeta_{2,3}$ are defined as:
\begin{align}
  \chi_{2,3} & = \frac{u_{0}}{M_{0}} \Big[ 1 + \imath \, \frac{u_{1}}{u_{0}} \Delta + \frac{M_{1}}{(M_{0})^{2}} \Delta \notag \\
  & \quad \quad {} + \big( \frac{u_{2}}{u_{0}} + \frac{u_{1} M_{1}}{2 u_{0} (M_{0})^{2}} - \frac{3 (M_{1})^{2}}{8  (M_{0})^{4}} - \frac{M_{2}}{2 (M_{0})^{2}} \big) \Delta^{2} \Big] \label{eq:B29}\\
\gamma_{2,3} & = \frac{v_{0}}{M_{0}} \Big[ 1 + \imath \, \frac{v_{1}}{v_{0}} \Delta + \frac{M_{1}}{(M_{0})^{2}} \Delta \notag \\
  & \quad \quad {} + \big( \frac{v_{2}}{v_{0}} + \frac{v_{1} M_{1}}{2 v_{0} (M_{0})^{2}} - \frac{3 (M_{1})^{2}}{8  (M_{0})^{4}} - \frac{M_{2}}{2 (M_{0})^{2}} \big) \Delta^{2}  \Big] \label{eq:B30}\\
\zeta_{2,3} & = \frac{1}{M_{0}} \Big[ 1 + \frac{M_{1}}{ (M_{0})^{2}} \Delta - \big( \frac{3 (M_{1})^{2}}{8 (M_{0})^{4}} +  \frac{M_{2}}{2 (M_{0})^{2}} \big) \Delta^{2} \Big] \label{eq:B31} 
\end{align}
given:
\begin{align}
    M_{0} & = \sqrt{1 + |u_{0}|^{2} + |v_{0}|^{2}} \label{eq:B32} \\
    M_{1} & = \text{Im}[u_{1} u_{0}^{*}] + \text{Im}[v_{1}v_{0}^{*}] \label{eq:B33} \\
    M_{2} & = |u_{1}|^{2} + |v_{1}|^{2} + 2 \text{Re}[u_{0} u_{2}^{*}] + 2 \text{Re}[v_{0} v_{2}^{*}] \label{eq:B34}
\end{align}
with
\begin{align}
    v_{0} & = \Big(\frac{b}{c}\Big) \Bigg[ \frac{\lambda_{2,3} +|c|^{2} (\delta_{1} - \epsilon_{0-1})}{\lambda_{2,3} - |b|^{2}\delta_{1}} \Bigg] \label{eq:B35} \\
    v_{1} & = \Big(\frac{b}{c}\Big) \Bigg[ \frac{1}{(\lambda_{2,3} - |b|^{2}\delta_{1})^{2}} \Bigg] \notag \\
    & \quad \times \Bigg[ \big( \lambda_{2,3} + |c|^{2}(\delta_{1} - \epsilon_{0-1}) \big) \big[\lambda_{2,3}\beta_{1-1} - |b|^{2} \delta_{3} \big]\notag \\
    & \quad \quad \quad + \big(\lambda_{2,3} \beta_{10} - |c|^{2} \delta_{2}\big) \big[|b|^{2}\delta_{1} - \lambda_{2,3}\big] \Bigg] \label{eq:B36} \\
    v_{2} & = \Big(\frac{b}{c}\Big) \Bigg[ \frac{1}{(\lambda_{2,3} - |b|^{2}\delta_{1})^{3}} \Bigg] \notag \\
    & \: \: \: \times \Bigg[ (\lambda_{2,3} - |b|^{2}\delta_{1}) (\lambda_{2,3}\beta_{1-1} - |b|^{2}\delta_{3}) (\lambda_{2,3}\beta_{10} - |c|^{2}\delta_{2}) \notag \\
    & \quad \: \: - (\lambda_{2,3} - |b|^{2}\delta_{1})^{2} (\lambda_{2,3}\eta^{*}_{10} - |c|^{2}\delta_{4}) \notag \\
    & \quad \: \:  - \Big[ (\lambda_{2,3} - |c|^{2}(\delta_{1} - \epsilon_{0-1})) \big[(\lambda_{2,3}\beta_{1-1} - |b|^{2}\delta_{3})^{2}  \notag \\
    & \quad \qquad - (\lambda_{2,3} - |b|^{2} \delta_{1}) (|b|^{2}\delta_{5} + \lambda_{2,3}\eta_{1-1}^{*} ) \big] \, \Big] \, \Bigg] \label{eq:B37}
\end{align}

\vspace{1cm}

and
\begin{align}
    u_{0} & = \Big(\frac{-1}{c a^{*}}\Big) \Big[ |c|^{2} + v_{0} c b^{*} \Big] \label{eq:B38}\\
    u_{1} & = \Big(\frac{-1}{c a^{*}}\Big) \Big[|c|^{2}\beta_{1-1} + c b^{*}\big(v_{1} + v_{0}(\beta_{1-1} - \beta_{0-1})\big) \, \Big] \label{eq:B39} \\
    u_{2} & = \Big(\frac{1}{c a^{*}}\Big) \Bigg[\lambda_{2,3} + |c|^{2}(\eta_{1-1} - \epsilon_{1-1}) \notag \\
    & \quad \quad \quad - c b^{*} \big[ v_{2} + v_{1}(\beta_{0-1} - \beta_{1-1}) \notag \\
    & \quad \quad \qquad + v_{0} (\beta_{0-1}\beta_{1-1} + \epsilon_{1-1} - \eta_{1-1} - \eta_{0-1}^{*}) \, \big] \, \Bigg] \label{eq:B40}
\end{align}

\vspace{1cm}

with the additional mixing parameters:
\begin{align}
    \delta_{2} & = \beta_{1-1}\eta_{0-1} - \beta_{0-1}\eta_{1-1} + \nu_{0-1} + \nu_{10}^{*} - \nu_{1-1}^{*} \label{eq:B41} \\
    \delta_{3} & = \beta_{10}\eta_{0-1}^{*} - \beta_{0-1}\eta_{10}^{*} - \nu_{0-1}^{*} - \nu_{10}^{*} + \nu_{1-1}^{*} \label{eq:B42} \\
    \delta_{4} & = \beta_{1-1}\nu_{0-1} - \beta_{0-1}\nu_{1-1}^{*} + \eta_{0-1}\eta_{1-1}^{*} \notag \\
    & \qquad \qquad \qquad \qquad \quad + \kappa_{0-1} - \kappa_{10}^{*} + \kappa_{1-1}^{*} \label{eq:B43} \\
    \delta_{5} & = \beta_{10}\nu_{0-1}^{*} + \beta_{0-1}\nu_{10}^{*} - \eta_{0-1}^{*}\eta_{10}^{*}\notag \\
    & \qquad \qquad \qquad \qquad \quad - \kappa_{0-1}^{*} - \kappa_{10}^{*} + \kappa_{1-1}^{*} \label{eq:B44}
\end{align}
\noindent where for $\ket{\psi_{2}}$ one chooses $\lambda_{2}$ defined in Eqn.~\ref{eq:B11}, with a similar convention for  $\ket{\psi_{3}}$.  

These solutions for the eigenstates exhibit the increased complexity present in the three dimensional case.  The numerical factors in Eqns.~\ref{eq:B22} - \ref{eq:B27} and Eqns.~\ref{eq:B35} - \ref{eq:B40} showcase a complicated interplay between the environmental factors $\beta_{ij}$, $\eta_{ij}$, $\epsilon_{ij}$, $\nu_{ij}$, and $\kappa_{ij}$ for the three different states. The time dependence of the top eigenstate, $\ket{\psi_{1}}$, is reminiscent of the two state solutions in that the leading order real time dependence is quadratic with a linear complex phase.  However, note that for general $H^{1}_{e}$, $H^{0}_{e}$, and $H^{-1}_{e}$ the leading order real time dependence for $\ket{\psi_{2}}$ and $\ket{\psi_{3}}$ is actually linear if one considers Eqns.~\ref{eq:B29} - \ref{eq:B31}. However, this linear time dependence will disappear if all the environmental factors $\beta_{ij}$, $\eta_{ij}$, $\epsilon_{ij}$, and $\nu_{ij}$ are purely real, due to the vanishing of Eqn.~\ref{eq:B33} in that limit.


In the case where the linear time dependence is present, one can think of it this way:  All aspects of the way in which the initial state is being diminished are occurring at a quadratic rate (to leading order).  But in the qutrit case, the probability is flowing from the initial state into a two dimensional subspace orthogonal to the initial state. Under certain conditions it is possible for the description of the system in this orthogonal subspace to move around with a linear time dependence, even as the profile of the evolution of the initial state remains quadratic.  

\begin{figure}[h!]
\centering
\includegraphics[width=3.5in]{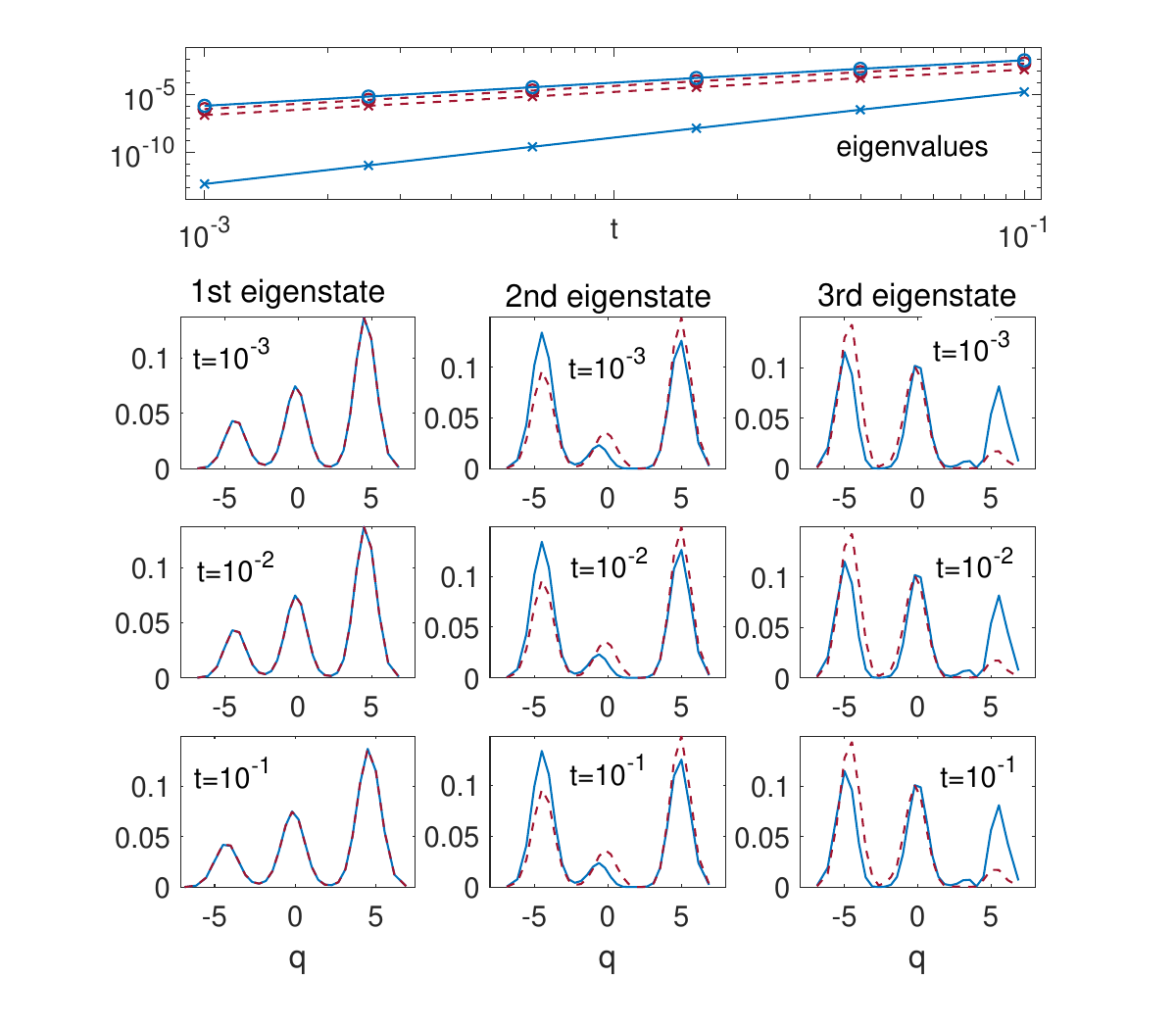}
\caption{The evolution of an initial Schr\"odinger cat state formed from three wavepackets. Eigenstates and the 2nd (circle markers) and 3rd (``x'' markers) eigenvalues of $\rho_s$ are shown in a similar manner to Fig.~\ref{fig:CopyCat}.  The Hamiltonians used are given by  the original ACL model (Eqn.~\ref{eq:5}, solid curves) and a modified ACL model (Eqn.~\ref{eqn:ACLfor3}, dashed curves).
These results allow us to link the qutrit results from this Appendix to the behaviors of more complex systems.  The quadratic (or faster) time dependence of the eigenvalues and transient stability of the eigenstates express the main features of the copycat process, even for these generalized cases.   Note that the solid and dashed curves for the first eigenstate overlap completely.
}
\label{fig:3packetcats}
\end{figure}

In this paper we took a phenomenon observed in the ACL model and provided a systematic analysis in the simpler case of the single qubit system, and in this Appendix we've extended the analysis to the qutrit. We now circle back to the more complicated ACL case. Figure~\ref{fig:3packetcats} shows results from calculations similar to the ACL calculations shown in Fig.~\ref{fig:CopyCat}, but with an initial state comprised of three wavepackets.  The broad features of the copycat process are also present in this more complicated case~\footnote{Figure~\ref{fig:3packetcats} shows a more narrow time range than Fig~\ref{fig:CopyCat} because we had to wait until $t\approx 10^{-3}$ for the third eigenvalue to resolve numerically, and the limited overall size of the system Hilbert space forced us to place the three packets too close together to einselect cleanly.  The latter restriction is irrelevant to the points we make here which are about early time behavior.}~\footnote{In principle one could compare the forms of the eigenstates in Fig.~\ref{fig:3packetcats} with the analytical results for the qutrit---as well as derive the qutrit versions of other quantities discussed in Secs.~\ref{sec:Implications} and~\ref{sec:Beyond}---but we did not feel such an involved analysis would add much of interest to this paper.}.

In our numerical explorations of different models we always found the second largest eigenvalue of $\rho_s$ evolved as $t^2$ at early times, and the smaller eigenvalues evolved as an even power of $t$ greater than or equal to two.  This faster than linear evolution of the eigenvalues, in comparison with the slower evolution of the eigenstates, is an essential part of the copycat process. To achieve the $t^2$ behavior for the third eigenvalue shown by dashed curve in Fig.~\ref{fig:3packetcats} (which matches our qutrit analysis) we used the ACL model with the interaction term modified to give
\begin{align} \label{eqn:ACLfor3}
 H_{w} & = H^{s}_{SHO} \otimes 1^{e} \notag \\
 & \quad \quad  + 
 \left( \sum_{i=1}^{10}\left|q_i\right> q_i \left<q_i\right| \right)\otimes H^{e}_A \notag \\
  & \quad \quad  +\left( \sum_{i=11}^{20}\left|q_i\right> q_i \left<q_i\right| \right)\otimes H^{e}_B \notag \\ 
  & \quad  \quad +\left( \sum_{i=21}^{30}\left|q_i\right> q_i \left<q_i\right| \right)\otimes H^{e}_C
 + H^{e} \otimes 1^{s},
 \end{align}
 where $H^e_A$, $H^e_B$ and $H^e_C$ are each independently generated random hermitian matrices and the three sums divide the eigenstates of $q$ (defined in the 30-dimensional system subspace) into three equal ranges. Note that Eqn.~\ref{eqn:ACLfor3} is a more direct generalization of Eqn.~\ref{eq:B2} (used for the qutrit) than the original ACL Hamiltonian (Eqn.~\ref{eq:5}). 
\clearpage

\bibliography{ccLib}

\end{document}